\documentclass[10pt,letterpaper,twocolumn]{article} 
\usepackage{color}
\usepackage{ol2}
\usepackage[draft]{hyperref}
\usepackage{amsmath}

\begin{document}

\twocolumn[ 

\title{From Classical Four-Wave Mixing to Parametric Fluorescence in Silicon micro-ring resonators}


\author{Stefano Azzini$^1$, Davide Grassani$^1$, Matteo Galli$^1$, Lucio Claudio Andreani$^1$, Marc Sorel$^2$, Michael J. Strain$^2$, L.G. Helt$^3$, J.E. Sipe$^3$, Marco Liscidini$^1$, Daniele Bajoni$^{4,*}$ }

\address{
$^1$Dipartimento di Fisica, Universit\`{a} degli Studi di Pavia, via Bassi 6, Pavia, Italy
\\
$^2$School of Engineering, University of Glasgow, Glasgow G12 8LT, UK \\
$^3$Department of Physics and Institute for Optical Sciences, University of Toronto, 60 St. George Street, Ontario, Canada \\
$^4$Dipartimento di Ingegneria Industriale e dell'Informazione, Universit\`{a} degli Studi di Pavia, via Ferrata 1, Pavia, Italy
\\
$^*$Corresponding author: daniele.bajoni@unipv.it
}

\begin{abstract}

Four-wave mixing can be stimulated or occur spontaneously.  The first process is intrinsically much stronger, and well understood through classical nonlinear optics. The latter, also known as parametric fluorescence, can be explained only in the framework of a quantum theory of light. We experimentally demonstrate that, in a micro-ring resonator, there exists a simple relation between the efficiencies of these two processes, which is independent of the nonlinearity and size of the ring. In particular we show that the average power generated by parametric fluorescence can be immediately estimated from a classical FWM experiment. These results suggest that classical nonlinear characterization of a photonic integrated structure can provide accurate information on its nonlinear quantum properties.

\end{abstract}


 ] 

Mirco-ring resonators have been investigated for more than a decade, with applications ranging from signal processing to optical sensing \cite{little98,lipson08}. These resonators are extremely appealing for on-chip integrated optics and already exploited in commercial devices, as they can be realized either in silicon \cite{lipson04}, silicon nitride \cite{levy10}, or hydex \cite{ferrera08, razzari10} following CMOS compatible processes. 

The large field intensity that can be obtained by constructive interference within the ring has inspired several theoretical proposals and experiments in the field of nonlinear optics, including nonlinear bistability \cite{lipson04}, parametric frequency conversion \cite{absil00, tuner08,ferrera08}, parametric fluorescence \cite{clemmen09}, and frequency comb generation \cite{levy10,razzari10}. In this regard, four-wave mixing (FWM) is  probably the most investigated process. This is a third-order nonlinear effect that can be viewed as the elastic scattering of two photons of a beam (pump), which results in the generation of two new photons at different frequencies (idler and signal) \cite{boyd_book}. This nonlinear effect can occur with (stimulated FWM) or without (spontaneous FWM) an input signal beam. Spontaneous FWM, also known as parametric fluorescence, is  particularly interesting, as it can be exploited for the generation of correlated photon-pairs \cite{helt10}, used to construct qubits in quantum information and quantum computation.

\begin{figure}[t]
\centerline{\includegraphics[width=\columnwidth]{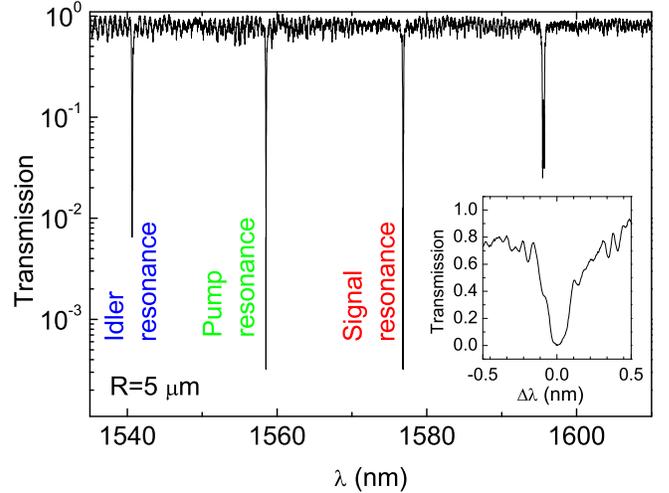}}
\caption{(Color online) Transmission spectrum of the $R=5$ $\mu$m ring. The inset shows a high resolution spectrum of the resonance at 1558.5.}
\label{Fig1}

\end{figure}

For a given device, there is obviously a connection between the efficiencies of spontaneous and classical FWM, as the two processes originate from the same material nonlinearlity. Thus, it is somehow expected that systems designed to enhance stimulated FWM can be utilized for the generation of correlated photon-pairs. Yet, it would be extremely helpful to know of the efficiency of spontaneous FWM in a device from nonlinear characterization via stimulated FWM \cite{helt12}. The stimulated process is indeed intrinsically much more efficient, and thus it can be observed much more easily. 

\begin{figure}[t]
\centerline{\includegraphics[width=\columnwidth]{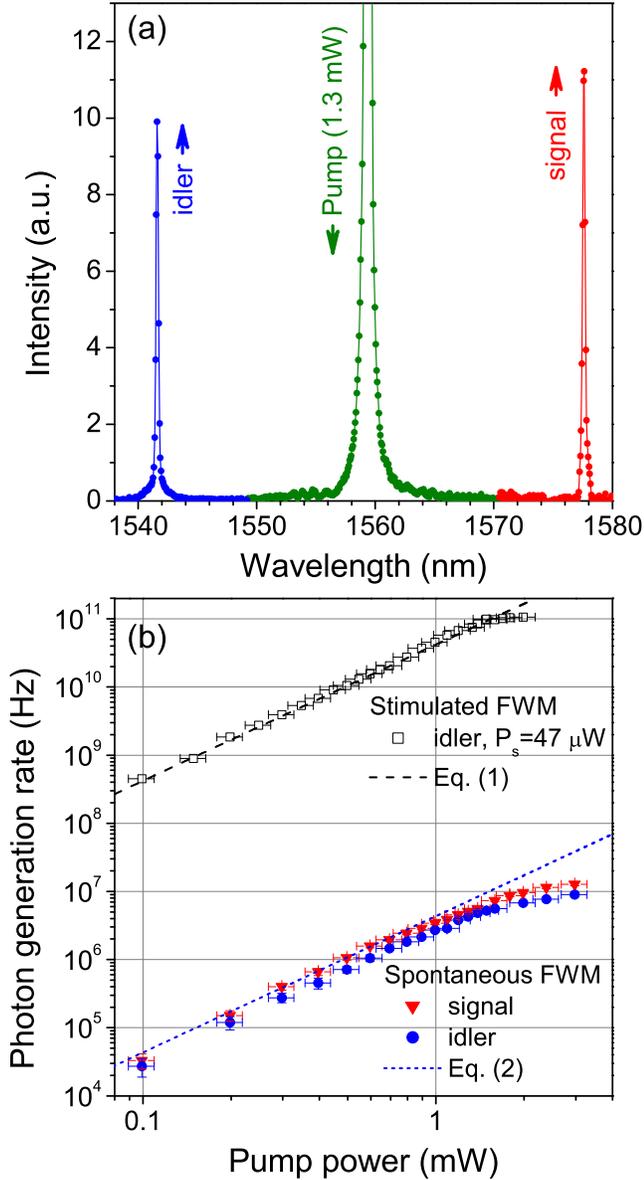}}
\caption{(Color Online) (a) An example of a spontaneous FWM spectrum for a R=5 $\mu$m ring resonator. (b) Scaling of the estimated number of generated signal (blue circles) and idler (red triangles) photons inside the ring for spontaneous FWM with varying $P_P$. Same for idler photons in classical FWM (black squares), with 47 $\mu$W injected at the signal resonance. The black dashed line is the best fit to the stimulated data from eq. (1), the blue short dash line is the theoretical prediction from eq. (2).}
\label{Fig2}
\end{figure}

In this work we study stimulated and spontaneous FWM in micro-ring resonators, and  we investigate their efficiencies with respect to the characteristic parameters of the ring: the radius R and the resonance quality factor Q. We consider single-channel side-coupled ring resonators. In this system, for any triplet of resonances equally separated in energy, assuming $\omega_s\simeq\omega_p\simeq\omega_i$, and considering a similar Q and group velocity $v_g$, the generated powers integrated over the idler resonance are 
\begin{equation}\label{stimulated}
P_{i,ST}=(\gamma 2\pi R)^2\left(\frac{Qv_g}{\omega_p\pi R}\right)^4P_sP_p^2.
\end{equation}
\begin{equation}\label{spontaneous}
P_{i,SP}=(\gamma 2\pi R)^2\left(\frac{Qv_g}{\omega_p\pi R}\right)^3\frac{\hbar\omega_p v_g}{4\pi R}P_p^2.
\end{equation}
for the stimulated and spontaneous processes, respectively \cite{note2}. These expressions can be calculated following  \cite{liscidini12},  assuming the critical coupling condition for the CW pump of power $P_p$, tuned to the central resonance, and  considering, in the stimulated case, a CW signal of power $P_s$  tuned to the lowest energy resonance. Here $\gamma$ is the nonlinear parameter of the ring waveguide \cite{helt10}.  It is interesting that, given the same input pump power $P_p$ the ratio between spontaneous and stimulated powers
\begin{equation}\label{ratio}
\frac{P_{i,SP}}{P_{i,ST}}=\frac{1}{4Q}\frac{\hbar\omega_p^2}{P_s}.
\end{equation}
is independent of the ring size, but it depends uniquely on the resonance quality factor, the signal power in the stimulated experiment, and a characteristic power  $\hbar\omega_p^2$ (e.g. at $\hbar\omega_p=0.8eV$, we have $\hbar\omega_p^2\simeq 160\mu W$). This makes it possible, given Q, to determine the number of pairs generated in the quantum process solely by means of the corresponding stimulated FWM experiment, once the value of Q is known.

The devices were fabricated on a silicon-on-insulator (SOI) wafer via e-beam lithography and inductively coupled plasma etching. Rings with radii of 5, 10, 20, and 30 $\mu$m were fabricated side-coupled to a single bus waveguide: the waveguides and rings have a 220$\times$500 nm$^2$ rectangular cross section. Spot-size converters are used for efficient coupling \cite{SSC2}. The silicon chip was finally coated with a protective PMMA layer. The samples were characterized using a swept laser: a transmission curve for R=5$\mu$m is shown in Fig. 1 (a). The spectrum shows well distinct resonances ($Q\sim7900$) with the transmission falling to less than 1\%, a sign that the ring is in critical coupling with the access waveguide. In the other samples the Q factors are $Q\sim8400$ for R=10$\mu$m,   $Q\sim12000$ for R=20$\mu$m and Q$\sim$15000 for R=30$\mu$m \cite{noteQ}.

For the nonlinear experiment a tunable laser (Santec TSL-510) set at the pump and, for the stimulated FWM experiments another set at the signal frequencies, were injected into the bus waveguide of the rings with a tapered lensed fiber. The emitted light was spectrally filtered and sent to a CCD detector. Fig. 2 shows the results of FWM experiments on the R=5$\mu$m ring resonator, the sample with the smallest footprint, using the resonances marked in Fig 1. The power coupled into the ring and that generated within the ring are estimated by measuring the losses of each component of the experimental setup and consequently rescaling the output power from the laser or the power measured by the CCD (which was calibrated before the experiment). Couplipg losses to/from the sample are obtained as half its total insertion loss (measured to be about 7 dB). We evaluate a total error of about 10\% on the estimated powers. Fig. 2 (a) shows an example of spontaneously generated signal and idler beams. The  widths of the peaks are identical to those measured in transmission experiments, indicating that there is little degradation due to free carrier absorption. 
The integrated intensities for the signal and idler beams for spontaneous FWM, as well as the integrated intensity of the idler beam for stimulated FWM at fixed signal input power ($P_s$=$200\mu$ W),  are shown in Fig. 2 (b) as a function of $P_p$. In all cases the intensities follow the expected quadratic dependence of \eqref{stimulated} and \eqref{spontaneous}. Notice that the saturation observed for $P_p>2$ mW is due to the thermo-optic effect induced by two-photon absorption, which produces a small redshift of the ring resonances \cite{lipson04,note}. We have also verified that in classic FWM the idler intensity scales linearly with $P_s$. 

\begin{figure}[t]
\centerline{\includegraphics[width=\columnwidth]{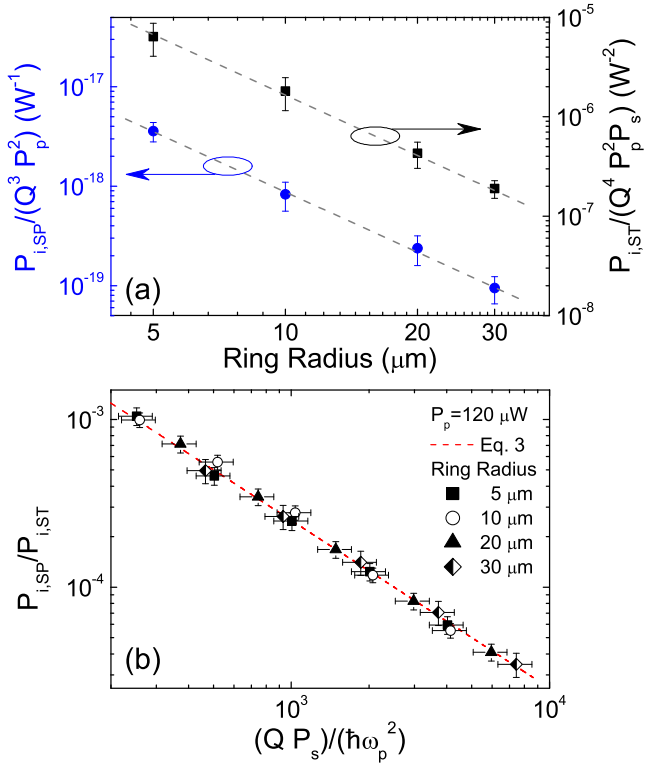}}
\caption{(Color Online) (a) Scaling of the generated idler intensities with the ring radius, for stimulated and spontaneous processes. The dashed lines are guides to the eye proportional to $R^{-2}$. (b)  Ratio between the generated idler photons in spontaneous and classical processes (see \eqref{ratio}) for all the four rings investigated. The dashed line is given by eq. \ref{ratio}. }
\label{Fig3}

\end{figure}

We performed a best fit of the stimulated FWM data (black dashed curve in Fig. 2 (b)) using $\gamma$ as the only fit parameter (we take $v_g=c/n_{eff}$ where $n_{eff}$=2.47 is estimated by numerical simulation and is a typical mode effective index for this kind of waveguides \cite{tuner08}), obtaining $\gamma$=190 W$^{-1}$m$^{-1}$, a value consistent with those already reported in \cite{tuner08}. We then used this value to compute the expected spontaneous emission rate using eq. (2). The result is shown as a short dashed blue line in Fig. 2(b) and is in very good agreement with the experiments. 
 
FWM experiments were carried out on rings with radii of 5, 10, 20 and 30 $\mu$m to verify the scaling of $\eqref{stimulated}$, \eqref{spontaneous}, and \eqref{ratio} with $R$ . In all cases, we chose rings near the critical coupling condition, and we worked with the same separation in energy between the signal, pump, and idler resonances: this means we have chosen the first neighbors in the case of the R=5$\mu$m ring (as in Fig 1(b)), second neighbors for R=10$\mu$m, fourth neighbors for R=20$\mu$m and sixth neighbors for R=30$\mu$m. In Fig. 3 (a) we show stimulated and spontaneous idler powers as a function of $R$: the data were averaged on a set of measures taken varying $P_p$ and $P_s$. Notice that, for all data of Fig. 3, we kept $P_p<1$ mW to avoid the thermo-optic effect. As expected from \eqref{stimulated} and \eqref{spontaneous}, both of the generated powers scale as $R^{-2}$. Finally, in Fig 3 (b) we show the ratio  between the measured generated idler powers in the spontaneous and in the classical processes as a function of $Q\cdot P_s$. The data show an excellent agreement with \eqref{ratio}  and confirm that the ratio is independent of the ring radius. More importantly, these results prove that data from stimulated FWM can be used to precisely predict spontaneous FWM generation rates.  For clarity, in Fig. 3(b) we show only data for $P_p$=0.12 mW, but we obtained the same results for all the investigated pump powers varying between 60 $\mu$W and 1 mW. 

In conclusion we have reported classical and spontaneous Four-wave mixing in silicon micro-ring resonators. We have verified that the efficiency of the spontaneous process is related to that of the classical process by a simple analytical formula. This result is of great practical interest, since it is general and not limited to ring resonators \cite{liscidini12}: given any photonic structure, the application of this method allows the derivation of the number of photon pairs generated in the spontaneous process through a simple measure of the classical process, greatly simplifying the characterization of any structure designed for the generation of quantum photonic states.

This work was supported by MIUR funding through the FIRB ``Futuro in Ricerca" project RBFR08XMVY and from the foundation Alma Mater Ticinensis.



\begin{thebibliography}{99}
\bibitem{note2} Unlike in Ref.[\onlinecite{helt12}], here we assume losses in the ring resonator. At the critical coupling,  the on-resonance  field enhancement in the ring resonator  is $FE\simeq\sqrt{\frac{2Qv_g}{\omega_0L}}$, with $\omega_0$ the resonant frequency (see also Ref.\cite{absil00}).
\bibitem{noteQ}
The variation of the quality factor between the signal, pump and idler resonances is below 2\% for all four rings. The value we have taken for the quality factor in the equations is an average between the Qs of the three resonances.
\bibitem{note} We verified that the quadratic trend is maintained up to the maximum available pump power if the pump wavelength is retuned to compensate for the thermo-optic redshift of the  resonances for $P_p> 2$ mW. This implies that Q degradation due to free carrier absorption can be neglected in this device for all the investigated $P_p$. The data of Fig. 2 are taken with a fixed value of the pump energy.
\end{thebibliography}
\end{document}